\documentclass{article}
\usepackage{amsmath,graphicx,mlspconf}
\usepackage{enumitem}
\setlist{nosep, leftmargin=14pt}
\let\svthefootnote\thefootnote
\newcommand\freefootnote[1]{%
  \let\thefootnote\relax%
  \footnotetext{#1}%
  \let\thefootnote\svthefootnote%
}
\usepackage[utf8]{inputenc}
\usepackage[english]{babel}
\usepackage{graphicx}
\usepackage{hyperref}
\usepackage{bm}
\usepackage{amsfonts} 
\usepackage{float}
\usepackage{xcolor}
\usepackage{amsmath}
\usepackage{caption}
\usepackage{subcaption}
\newcommand{\cmmnt}[1]{\ignorespaces}
\usepackage{eqnarray,amsmath}
\usepackage[linesnumbered,ruled,vlined]{algorithm2e}
\title{A statistical framework for model-based inverse problems in ultrasound elastography }
\name{%
    Narges Mohammadi$^{\star}$
    \qquad Marvin M. Doyley$^{\star}$%
    \qquad Mujdat Cetin$^{\star}{}^{\dagger}$
}
\address{$^{\star}$ Department of Electrical and Computer Engineering, University of Rochester, Rochester, NY, USA \\%
    $^{\dagger}$ Goergen Institute for Data Science, 
University of Rochester, Rochester, NY, USA
}

\begin{document}
%
\maketitle
\begin{abstract}
Model-based computational elasticity imaging of tissues can be posed as solving an inverse problem over finite elements spanning the displacement image. As most existing quasi-static elastography methods count on deterministic formulations of the forward model resulting in a constrained optimization problem, the impact of displacement observation errors has not been well addressed. To this end, we propose a new statistical technique that leads to a unified optimization problem for elasticity imaging. Our statistical model takes the imperfect nature of the displacement measurements into account, and leads to an observation model for the Young's modulus that involves signal dependent colored noise. To solve the resulting regularized optimization problem, we propose a fixed-point algorithm that leverages proximal splitting methods. Preliminary qualitative and quantitative results demonstrate the effectiveness and robustness of the proposed methodology.
\end{abstract}
\begin{keywords}
ultrasound elastography, computational imaging, elasticity imaging, Young’s Modulus, statistical modeling, proximal splitting methods. 
\end{keywords}
\section{Introduction}
Elasticity imaging is concerned with the problem of reconstructing tissue physical parameters in terms of elastic modulus distribution and is usually performed by techniques mainly adapted from deformation imaging and shear waves \cite{quasi5}. Knowledge of Young's modulus and shear modulus, broadly referred to as elasticity parameters, has a large number of applications in non-invasive medical diagnosis and tissue property characterization \cite{quasi4}, \cite{quasi1}\cmmnt{ as quantified behaviour of tissue}. \cmmnt{ Typically, any standard medical imaging method can be employed to measure the deformation behaviour\cmmnt{pattern}; although, to have high accuracy detection for tiny deformation, phase sensitivity ones as Ultrasond, MRI and optical imaging \cmmnt{\cite{insana}}\cite{quasi2} are more efficient for utilization.} 
In many ultrasound elastography problems, quasi-static loading is applied on the tissue which involves small force indentation \cmmnt{stimulus} moving slowly on the exterior surface and the overall interior tissue reacting to it robustly through the stationary state \cite{quasi5}\cmmnt{ \cite{quasi1}, }. \\
For estimating deformation fields, speckle tracking is generally utilized which relies on B-mode ultrasound images acquisition, before and after applying quasi-static force loading on the surface \cmmnt{loading (indentation)} \cite{quasi5},\cite{quasi2}. Moreover, to represent the tissue deformation, finite element methods (FEM) coupled with a generated mesh \cmmnt{\cite{mesh}} are exploited which seek to reflect the spatial distribution of force-displacement relationships as a function of the material geometry and boundary conditions \cite{book}. 
For the modulus elastography objective, vast majority of solvers for elasticity inverse problems are developed for the linear elastic tissues, \cmmnt{(specimen)} which required the use of a constitutive model, equilibrium equation or global stiffness equation as a forward model. In this regard, existing methods can be classified according to their adopted priors while majority of them can be characterized as model-based inverse methods \cite{marvin}, \cite{quasi4}.
\\
As reported in \cite{unique}, no unique solution is available for the quasi-static elastic problem from a single observation; accordingly, many approaches employing a variety of priors or regularizers have been developed for solving this ill-posed problem \cite{marvin}\cmmnt{,\cite{unique}}. 
\cmmnt{\cite{adjoint1}}\cmmnt{, \cite{adjoint2}}In \cite{adjoint3}, the authors propose approaches based on the adjoint method which require solving the system of three prior models iteratively utilizing gradient-based methods.
The methodology introduced in \cite{surface} considers the problem of elasticity imaging using surface displacement during multiple observations. \cmmnt{; however, multiple realization of force deformation might not be accessible in clinical applications.}
Recently, a Matlab package for quasi-static elasticity imaging has been developed in \cite{matlab} which attempts to solve the regularized least squares inverse problem involving a displacement correlation data fidelity term by Gauss-Newton method. This deterministic approach \cmmnt{which is based on perturbation method \cite{peturb}, }requires forward model solving for displacement estimation at each iteration and Jacobian computation of the global stiffness as a function of unknown elasticity parameters which introduces significant computation time for large number of nodes and exhibits solution instability and inaccuracy especially due to the effect of noise in Jacobian computation.
\\
To address some of the weaknesses of existing methods for ultrasound elastography, we propose a new statistical framework to reconstruct Young's elasticity modulus directly from a single measured displacement field and force boundary conditions. In this approach, a unified objective function is introduced by integrating the elasticity forward model and displacement realization model into the data fidelity term which involves an effective signal dependent correlated noise model. Moreover, the proposed inverse problem solution is developed by incorporating a total variation regularizer for the Young's modulus, which can be interpreted as leading to a Bayesian estimation problem and is efficiently solved using fixed-point methods and proximal splitting algorithms. Our preliminary results demonstrate the effectiveness of the proposed approach.
\\
The remainder of this article is organized as follows. We introduce the inverse problem of quasi-static elasticity imaging in Section \ref{backward}. The proposed statistical approach for solving the inverse elasticity problem is detailed \cmmnt{developed} in Section \ref{opt}. Simulations and Young's modulus reconstruction results are presented in Section \ref{sim}, and finally, Section \ref{conclude} contains some concluding remarks and identifies future work directions.
\section{Formulation of inverse problem for Quasi-static Elastography}
\label{backward}
Model-based elastography problems involve discretized forward models to express the force-displacement data relationship with the physical material property.
Typical ultrasound realizations consist of displacement fields $\mathbf{u}$ over all the mesh nodes and applied force fields $\mathbf{f}$ over the surface. Let the number of mesh nodes be denoted by $\mathbf{N}$ which is the same as the number of Elastic parameters $\mathbf{E}$. In this setting, the ideal forward model in the context of the global stiffness equation can be described as:
\begin{equation}
\label{eq:1}
\mathbf{f}_{true}=\mathbf{K}(\mathbf{E})\mathbf{u}
\end{equation}
where $\mathbf{f}_{true}\in \mathbb{R}^{2N\times 1}$ indicates the global nodal force vector constructed by the noiseless true force boundary conditions in lateral and axial directions (which leads to the dimension of $2N$) and $\mathbf{u}\in \mathbb{R}^{2N\times 1}$ represents the global nodal displacement vector. Moreover, the global stiffness matrix $\mathbf{K}(\mathbf{E})\in  \mathbb{R}^{2N\times 2N}$, which is a function of Young's elasticity modulus $\mathbf{E}\in \mathbb{R}^{N\times 1}$, describes the stiffness behavior of the entire mesh structure over the nodes.\\
The inverse problem of finding the spatial distribution of Young's modulus of elasticity can be turned to the problem of solving a constrained optimization problem. Typical approaches to solve this problem require the minimization of objective functions of the following form, based on regularized displacement error, subject to a constraint involving the forward model:
\begin{equation}
\label{eq:2}
\begin{array}{l}
\mathrm{argmin} _{\mathbf{u},\mathbf{E}} \left \| \mathbf{u}-\mathbf{u^{m}} \right \|_{2}^{2}+\lambda \|\nabla\mathbf{E}\|_{1}\\
\\\quad\quad
s.t. \quad \mathbf{K}(\mathbf{E})\mathbf{u}-\mathbf{f}_{true}=0
\end{array}
\end{equation}
where $\mathbf{u^{m}}\in \mathbb{R}^{2N\times 1}$ describes the measured displacement vector consisting of measured lateral and axial displacements for all nodes, $\lambda$ is the regularization parameter, and  $\|\nabla\mathbf{E}\|_{1}$ is the total variation (TV) regularization term which is an appropriate choice for image reconstruction that promotes image smoothness while preserving edges\cmmnt{\cite{tvmarvin},\cite{tv} }.\\
This problem was solved previously as a deterministic inverse problem in \cite{marvin3} by Gaussian-based perturbation method
which iteratively updates the true displacements by forward model enforcement ($\mathbf{K}(\mathbf{E})\mathbf{u}-\mathbf{f}_{true}=0$) and updates the elasticity modulus by minimizing the regularized displacement error ($ \left \| \mathbf{u}-\mathbf{u^{m}} \right \|_{2}^{2}+\lambda \|\nabla\mathbf{E}\|_{1}$) using a perturbation method \cite{marvin3}. As this approach requires partial derivative matrix computation, it is highly sensitive to noise, addressing of which requires analytical noise modeling. In this regard, we pursue a statistical approach starting with the following observation model:
\begin{equation}
\label{eq:3}
\mathbf{f}=\mathbf{K}(\mathbf{E})\mathbf{u}+\mathbf{w}\qquad \mathbf{w}\sim \mathcal{N}(0,\,\bm{\Sigma_{w}})
\end{equation}
where $\mathbf{f}\in \mathbb{R}^{2N\times 1}$ is the force vector containing the measured force boundary conditions and $\mathbf{w}\in \mathbb{R}^{2N\times 1}$ is the nodal Gaussian noise vector. 
We can relate the measured displacements $\mathbf{u^{m}}$, commonly acquired by spectral tracking, to the underlying true displacements $\mathbf{u}$ by:
\begin{equation}
\label{eq:4}
\mathbf{u^{m}}=\mathbf{u}+\mathbf{n}\qquad \mathbf{n}\sim \mathcal{N}(0,\,\bm{\Sigma_{n}})
\end{equation}
where $\mathbf{n}\in \mathbb{R}^{2N\times 1}$ is the displacement noise vector with covariance $\bm{\Sigma_{n}}$ which features distinct noise variance in lateral and axial directions.\\
For estimating the Young's modulus $\mathbf{E}$ in this statistical framework, we extract the unknown elasticity parameters from the global stiffness matrix as an explicit vector by rearranging the forward model. In this regard, we benefit from the matrix $\mathbf{D}(\mathbf{u})\in \mathbb{R}^{2N\times N}$ \cite{FEM} which has the following relationship with $\mathbf{K}(\mathbf{E})$ using a 3D tensor $\bm{\Psi}\in \mathbb{R}^{N\times 2N\times2N}$ constructed by the equilibrium equation, Neumann boundary conditions, Poisson's ratio $\nu$, and node-element conversions:
\begin{equation}
\label{eq:5}
\mathbf{D}(\mathbf{u})\mathbf{E}=\mathbf{K}(\mathbf{E})\mathbf{u}\\
\end{equation}
\begin{equation}
\label{eq:6}
\mathbf{D}(\mathbf{u})=(\bm{\Psi}\mathbf{u})^{T}
\end{equation}
\begin{equation}
\label{eq:7}
\mathbf{K}(\mathbf{E})=\bm{\Psi}^{T}\mathbf{E}
\end{equation}

By applying these formulations to (\ref{eq:3}) and considering statistical modeling of observations error, we seek to solve an integrated linear-algebraic optimization problem with respect to the unknown elasticity modulus $\mathbf{E}$ without the need to explicitly estimate the displacement vector $\mathbf{u}$.
\section{Optimization Problem Formulation}
\label{opt}
Integrating the statistical observation model in (\ref{eq:3}) with the displacement realization process introduced in (\ref{eq:4}), leads to: 
\begin{eqnarray}
\label{eq:8}
\mathbf{f}&=&\mathbf{K}(\mathbf{E})\mathbf{u}+\mathbf{w}=\mathbf{K}(\mathbf{E})(\mathbf{u^{m}}-\mathbf{n})+\mathbf{w}\nonumber\\
&=&\mathbf{K}(\mathbf{E})\mathbf{u^{m}}-\mathbf{K}(\mathbf{E})\mathbf{n}+\mathbf{w}
\end{eqnarray}
Letting ${\mathbf{\Tilde{w}}}=-\mathbf{K}(\mathbf{E})\mathbf{n}+\mathbf{w}$ and using the noisy form of (\ref{eq:5}) $\mathbf{D}(\mathbf{u^{m}})\mathbf{E}=\mathbf{K}(\mathbf{E})\mathbf{u^{m}}$, the overall observation model could be represented as:
\begin{equation}
\label{eq:9}
\mathbf{f}=\mathbf{D}(\mathbf{u^{m}})\mathbf{E}+\mathbf{\Tilde{w}}\qquad \mathbf{\Tilde{w}}\sim \mathcal{N}(0,\,\bm{\Gamma})
\end{equation}
where $\bm{\Gamma}$ is given by:
\begin{equation}
\label{eq:10}
\bm{\Gamma}=\bm{\Sigma_{w}}+\mathbf{K}(\mathbf{E})\bm{\Sigma_{n}}\mathbf{K}(\mathbf{E})^{T}
\end{equation}
Given observations $\mathbf{f}$ and $\mathbf{u^{m}}$, the problem is to estimate $\mathbf{E}$. One way to interpret (\ref{eq:9}) is as a measurement model involving signal-dependent colored noise $\mathbf{\Tilde{w}}$. To estimate $\mathbf{E}$, we construct a TV regularized cost function, which is equivalent to a {\it  maximum a posteriori} (MAP) estimation formulation with a Laplacian prior distribution for $\mathbf{E}$: 
\begin{equation}
\label{eq:11}
\begin{array}{l}
\mathbf{\hat{E}}=\mathrm{argmin} _{\mathbf{E}}\quad\frac{1}{2}\left \|  \mathbf{f}-\mathbf{D}(\mathbf{u^{m}})\mathbf{E} \right \|_{{\bm{\Gamma}}^{-1}}^{2}+\frac{N}{2}\mathrm{log}\left | \bm{\Gamma} \right |+\lambda \|\nabla\mathbf{E}\|_{1}\\
\quad\quad\quad
s.t.\quad \mathbf{E}>0
\end{array}
\end{equation}
where $\left \| \mathbf{A} \right \|_{\mathbf{B}}^{2}:=(\mathbf{A}^{T}\mathbf{B}\mathbf{A})$. We use a fixed-point approach to solve (\ref{eq:11}), where $\bm{\Gamma}$ will be fixed when we take update $\mathbf{E}$, and then $\bm{\Gamma}$ will be updated with the new $\mathbf{E}$ based on (\ref{eq:10}).
For updating $\mathbf{E}$ in each step of the fixed-point approach, we leverage proximal splitting methods \cite{proximal}:
\begin{equation}
\label{eq:14}
\mathbf{E}_{n+1}=\textrm{prox}_{\mathbf{E}_{n}>0}(\textrm{prox}_{\gamma _{n}TV}(\mathbf{E}_{n}-\gamma _{n}\nabla g(\mathbf{E}_{n})))
\end{equation}
where:
\begin{equation}
\label{eq:15}
g(\mathbf{E})=\frac{1}{2}(\mathbf{f}-\mathbf{D}(\mathbf{u^{m}})\mathbf{E}  )^{T}\bm{\Gamma}^{-1}(\mathbf{f}-\mathbf{D}(\mathbf{u^{m}})\mathbf{E}  )
\end{equation}
\begin{equation}
\label{eq:16}
\nabla g(\mathbf{E})=-(\mathbf{D}(\mathbf{u^{m}}))^{T}\bm{\Gamma}^{-1}(\mathbf{f}-\mathbf{D}(\mathbf{u^{m}})\mathbf{E} )
\end{equation}
It is worth mentioning that favoring the power of proximal splitting methods, various types of regularizers could be employed as the split terms of the objective function are used distinctly during the optimization procedure. The Pseudo-code of the proposed approach is provided in \textbf{Algorithm 1}.
\section{Simulation Results}
\label{sim}
To evaluate the performance of the proposed approach for elasticity imaging, we attempt to reconstruct Young's modulus $\mathbf{E}$ from noisy synthetic measurements of force $\mathbf{f}$ and displacement field $\mathbf{u^{m}}$.
\begin{algorithm}[t]
\DontPrintSemicolon
\SetAlgoLined
\KwIn{$\mathbf{f}$, $\mathbf{u^{m}}$, $\lambda$, $Iter_{1}$, $Iter_{2}$, $\hat{\mathbf{E}}_{0}$}
\KwOut{Young's modulus field $\hat{\mathbf{E}}$}
 \For{$j=0,...,Iter_{1}-1$}{
   $\bm{\Gamma}_{j}=(\bm{\Sigma_{w}}+\mathbf{K}(\hat{\mathbf{E}}_{j})\bm{\Sigma_{n}}\mathbf{K}(\hat{\mathbf{E}}_{j})^{T}) \hfill\textrm{\textcolor{blue}{//Correlated covariance update}}$\;
    
    $\mathbf{E}_{0}$=$\hat{\mathbf{E}}_{j}$\;
      \For {$k=0,...,Iter_{2}-1$}{
            $\nabla g(\mathbf{E}_{k})=-(\mathbf{D}(\mathbf{u^{m}}))^{T}\bm{\Gamma}_{j}^{-1}(\mathbf{f}-\mathbf{D}(\mathbf{u^{m}})\mathbf{E}_{k} )$\;
        $\mathbf{E}_{k+1}=\textrm{prox}_{\mathbf{E}_{k}>0}(\textrm{prox}_{\gamma _{k}TV}(\mathbf{E}_{k}-\gamma _{k}\nabla g(\mathbf{E}_{k}))) \hfill\textrm{\textcolor{blue}{//Young's modulus update}}$\;

} \Return{$\mathbf{E}_{Iter2}$}\;
$\hat{\mathbf{E}}_{j+1}=\mathbf{E}_{Iter2}$;}
\Return{$\hat{\mathbf{E}}_{Iter1}$}\;
\Return{$\hat{\mathbf{E}}=\hat{\mathbf{E}}_{Iter1}$}\;
 \caption{Computational Elasticity Imaging Procedure}
\label{Alg1}
\end{algorithm}

 It should be mentioned that force fields $\mathbf{f}$ are obtained using surface displacements and boundary conditions based on applied loading force. 
Leveraging FEM, an irregular mesh composed of triangle elements is employed for discretization of all parameters over nodes of the mesh. 
Then, synthetic displacement fields $\mathbf{u}$ are calculated by solving the ideal forward problem ($\mathbf{K}(\mathbf{E})\mathbf{u}-\mathbf{f}_{true}=0$) for a specified distribution of material property $\mathbf{E}$ corresponding to a circle inclusion of 50KPa elasticity modulus  with specified radius and geometric location in a homogeneous background with 10KPa modulus elastogram. 
Multivariate Gaussian noise with covariance $\bm{\Sigma_{n}}$ is added to these displacement fields to synthetically generate noisy displacement fields $\mathbf{u^{m}}$. The noise level $\Delta =\left \| \mathbf{u^{m}}-\mathbf{u} \right \| / \left \| \mathbf{u^{m}} \right \|$ \cite{adjoint1} is set to 9\% and 3\% in the lateral and axial directions, respectively, leading to an overall SNR of 25 dB. \cmmnt{By aforementioned assumptions of noise level, the SNR of displacement measurements would be computed as in range of $20-30 \mathrm{dB}$. }Poisson's ratio $\nu$ is set to 0.495 and Neumann Boundary conditions are applied to the top and bottom boundaries of the tissue. Furthermore, Gaussian noise $\mathbf{w}$ is added to force measurements.
The step size $\gamma$ in (\ref{eq:14})  for optimal convergence is computed by the Lipschitz constant of the gradient of cost functions \cite{proximal}. The Python code of this implementation is available at GitHub \footnote{\url{https://github.com/narges-mhm/comp-elast}}.

\begin{figure}[t]
\begin{minipage}[b]{.48\linewidth}
  \centering
  \centerline{\includegraphics[width=3.5cm]{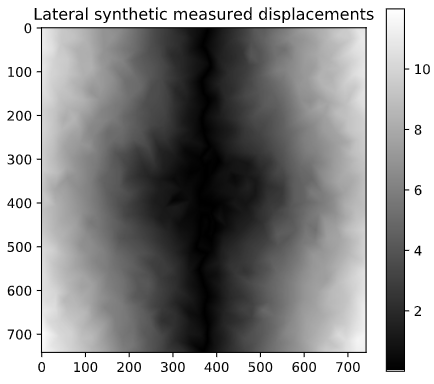}}
  \centerline{(a) \scriptsize{Lateral component of  $\mathbf{u}^{m}$}}\medskip
\end{minipage}
\hfill
\begin{minipage}[b]{0.48\linewidth}
  \centering
  \centerline{\includegraphics[width=3.5cm]{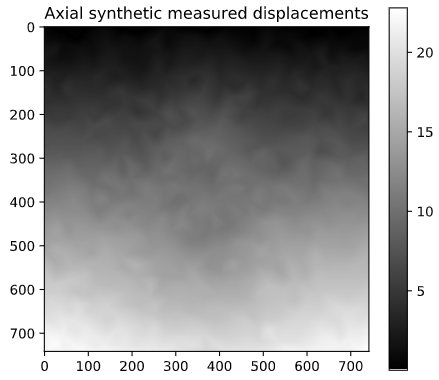}}
  \centerline{(b) \scriptsize{Axial component of $\mathbf{u}^{m}$}}\medskip
\end{minipage}
\begin{minipage}[b]{0.48\linewidth}
  \centering
  \centerline{\includegraphics[width=3.6cm]{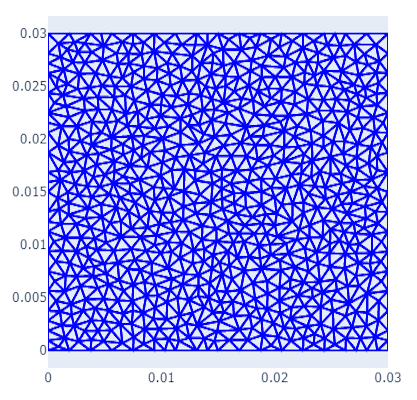}}
  \centerline{(c) \scriptsize{Irregular triangle mesh}}\medskip
\end{minipage}
\hfill
\begin{minipage}[b]{0.48\linewidth}
  \centering
  \centerline{\includegraphics[width=3.5cm]{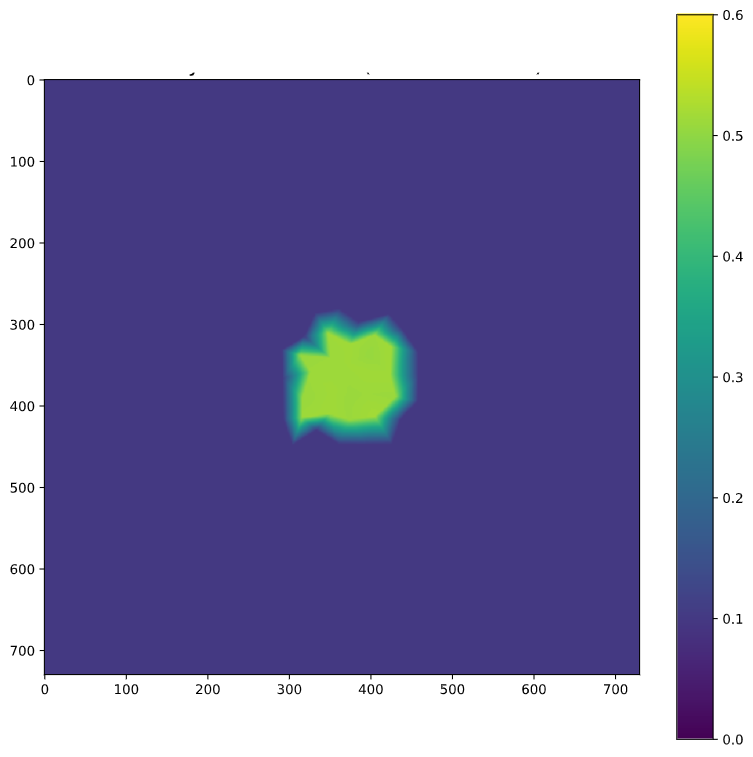}}
  \centerline{(d) \scriptsize{Reconstructed modulus $\hat{\mathbf{E}}$}}\medskip
\end{minipage}
\caption{Several components involved in our ultrasound elasticity imaging process. (a)-(b) Synthetic noisy displacements in lateral and axial directions. The color bars show displacements in $\mu m$. (c) Irregular triangle mesh used for medium cross-section discretization. (d) Reconstructed Young's modulus of elasticity $\hat{\mathbf{E}}$. The unit of the color bar is 100 KPa.}
\label{fig:fig1}
\end{figure}
Fig.~\ref{fig:fig1} illustrates several components involved in our elasticity imaging process. Fig.~\ref{fig:fig1}(a)-(b)
show the noisy displacement field measurements, which in practice could be obtained using speckle tracking algorithms on B-mode ultrasound images. For cross-section discretization of the domain, an irregular triangle mesh is utilized as shown in Fig.~\ref{fig:fig1}(c), which could be adaptively adjusted by measured displacement images. The Young's modulus image estimated by our algorithm is shown in Fig.~\ref{fig:fig1}(d). 

\begin{figure}[t]
\begin{minipage}[b]{0.48\linewidth}
  \centering
  \centerline{\includegraphics[width=3.5cm]{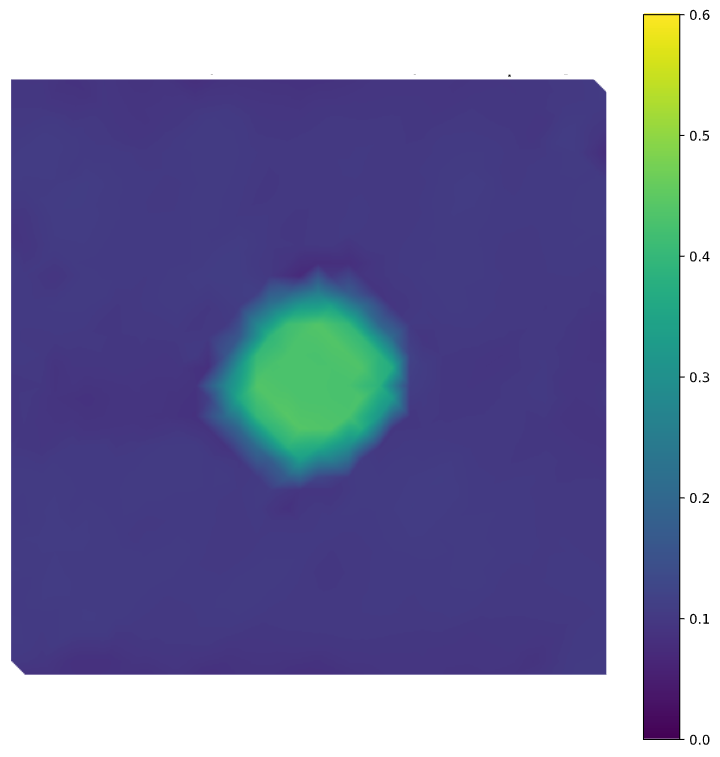}}
  \centerline{(a) \scriptsize{OpenQSEI (SNR=48dB).}}\medskip 
\end{minipage}
\hfill
\begin{minipage}[b]{0.48\linewidth}
  \centering
  \centerline{\includegraphics[width=3.5cm]{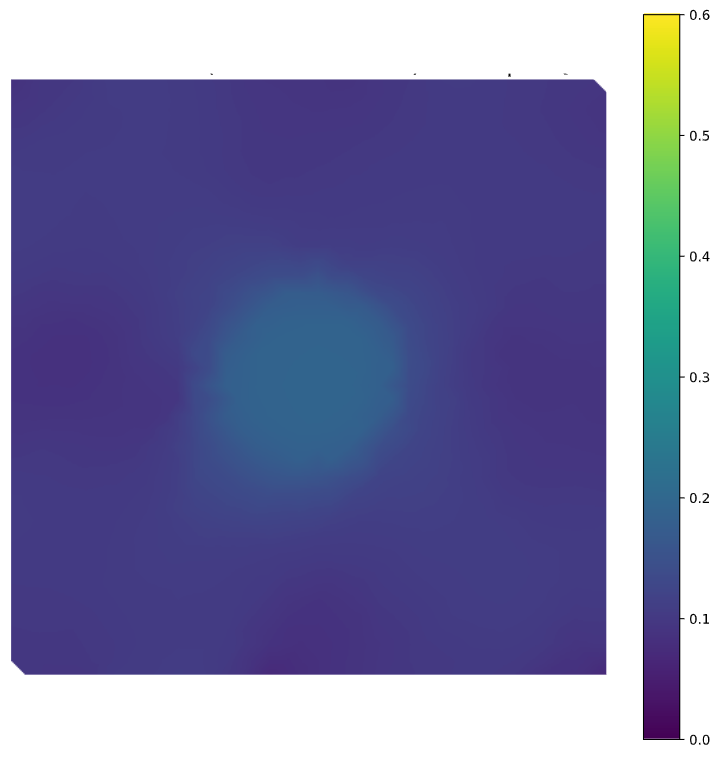}}
  \centerline{(b) \scriptsize{ OpenQSEI (SNR=25dB).}}\medskip 
\end{minipage}
\begin{minipage}[b]{0.48\linewidth}
  \centering
  \centerline{\includegraphics[width=3.5cm]{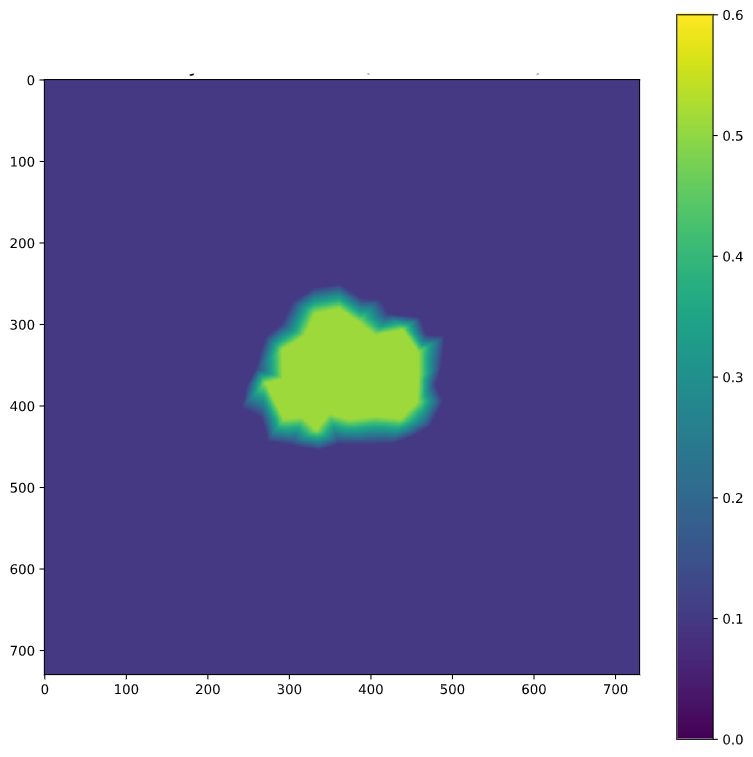}}
  \centerline{(c) \scriptsize{Propsoed (SNR=48dB).}}\medskip 
\end{minipage}
\hfill
\begin{minipage}[b]{0.48\linewidth}
  \centering
  \centerline{\includegraphics[width=3.5cm]{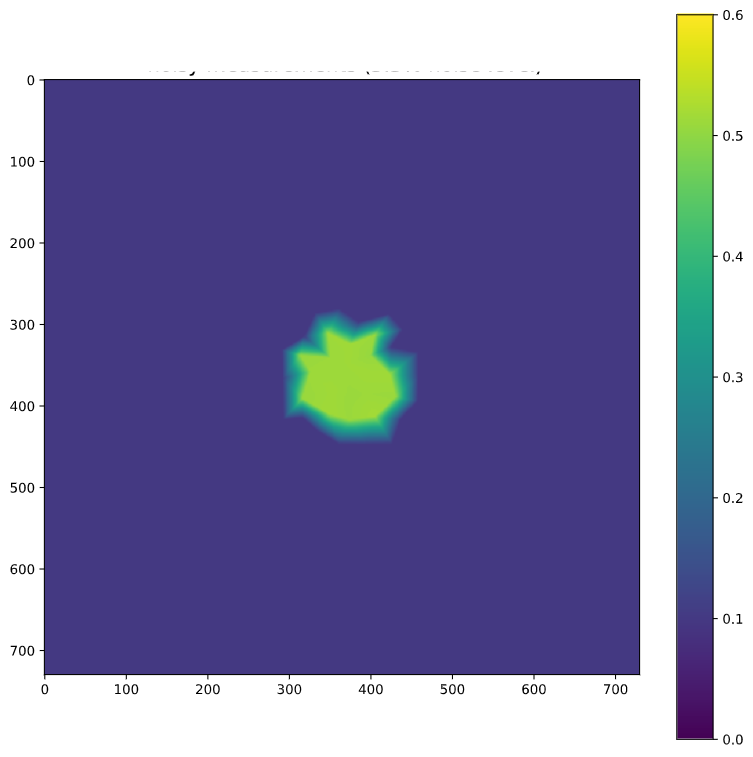}}
  \centerline{(d) \scriptsize{Proposed (SNR=25dB).}}\medskip 
\end{minipage}

\caption{Reconstructed Young's modulus $\hat{\mathbf{E}}$ images for two different levels of noise on measured displacements. (a)-(b) OpenQSEI. (c)-(d) Proposed approach. Both approaches use TV regularization. The unit of the color bar is 100 KPa.}
\label{fig:fig2}
\end{figure}

\begin{figure}[h]
\begin{minipage}[b]{.49\linewidth}
  \centering
  \centerline{\includegraphics[width=4.3cm]{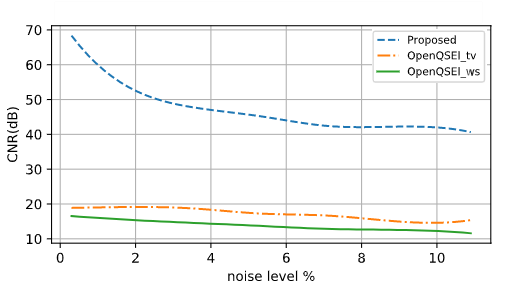}}
\end{minipage}
\vspace{2mm}
\begin{minipage}[b]{0.49\linewidth}
  \centering
  \centerline{\includegraphics[width=4.3cm]{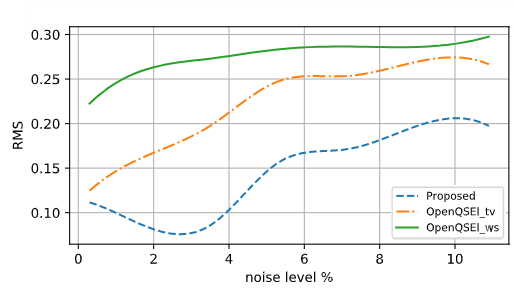}}
\end{minipage}
\caption{CNR and RMS performance metrics for noise levels $\Delta=0.1-10\%$ achieved by the proposed approach and OpenQSEI with TV and weighted-smoothness (ws) regularizers.}
\label{fig:fig3}
\end{figure}

\begin{figure}[t]
\begin{minipage}[b]{0.48\linewidth}
  \centering
  \centerline{\includegraphics[width=3.5cm]{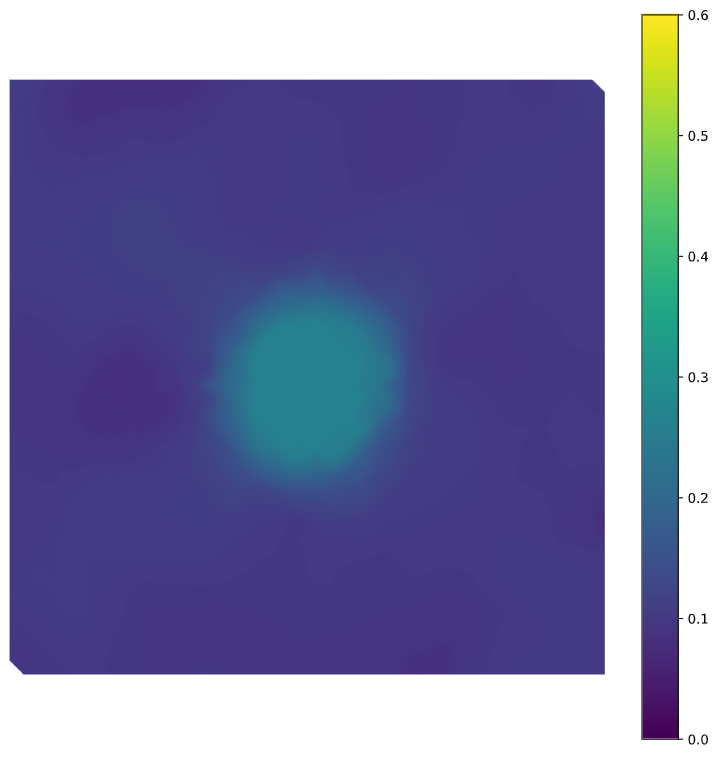}}
  \centerline{(a) \scriptsize{OpenQSEI ($\mathbf{E}_{true}=0.5$). }}\medskip
\end{minipage}
\hfill
\begin{minipage}[b]{0.48\linewidth}
  \centering
  \centerline{\includegraphics[width=3.5cm]{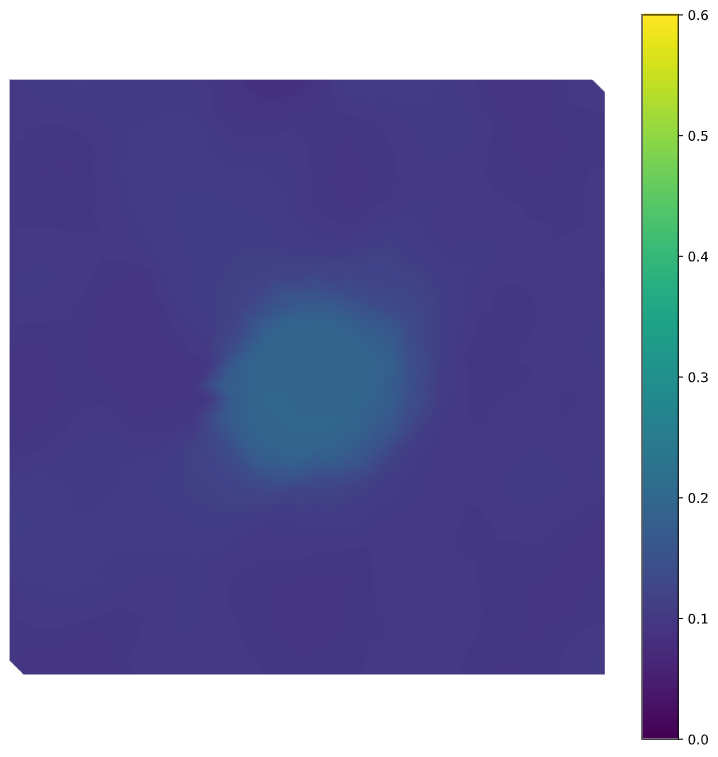}}
  \centerline{(b) \scriptsize{OpenQSEI ($\mathbf{E}_{true}=0.3$). }}\medskip
\end{minipage}
\begin{minipage}[b]{0.48\linewidth}
  \centering
  \centerline{\includegraphics[width=3.5cm]{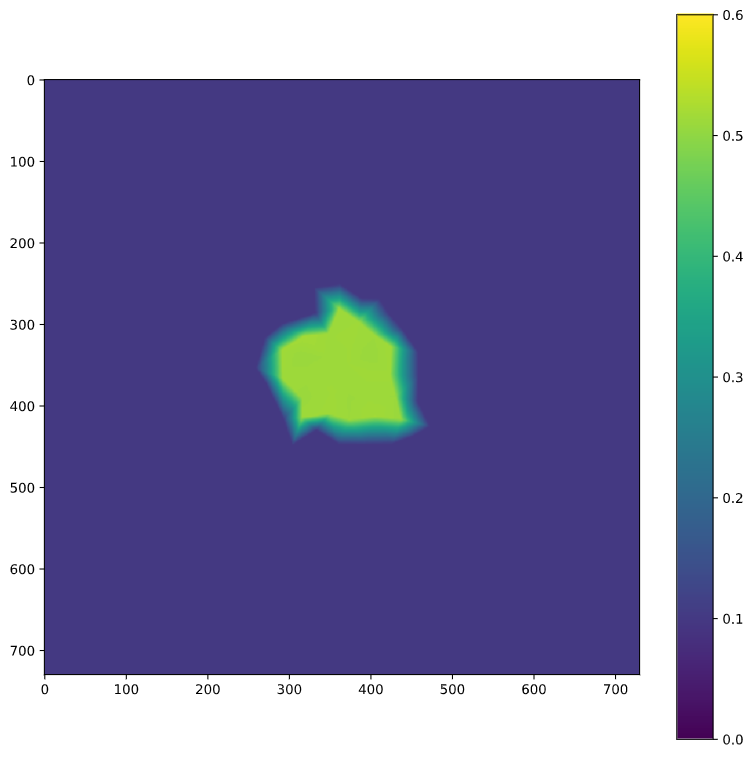}}
  \centerline{(c) \scriptsize{Proposed ($\mathbf{E}_{true}=0.5$).}}\medskip
\end{minipage}
\hfill
\begin{minipage}[b]{0.48\linewidth}
  \centering
  \centerline{\includegraphics[width=3.5cm]{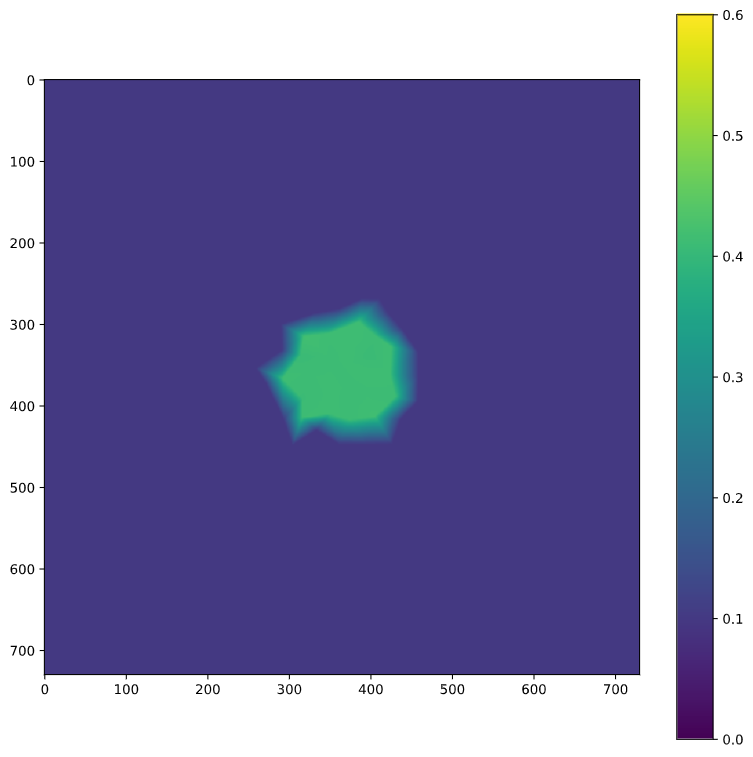}}
  \centerline{(d) \scriptsize{Proposed ($\mathbf{E}_{true}=0.3$).}}\medskip
\end{minipage}
\caption{Reconstructed Young's modulus $\hat{\mathbf{E}}$ images for two different settings of $\mathbf{E}_{true}$: 0.3 and 0.5 ($\times$100KPa). (a)-(b) OpenQSEI. (c)-(d) Proposed approach. Both approaches use TV regularization. SNR=30dB. The unit of the color bar is 100 KPa.}
\label{fig:fig4}
\end{figure}

\begin{figure}[h]
\begin{minipage}[b]{.49\linewidth}
  \centering
  \centerline{\includegraphics[width=4.3cm]{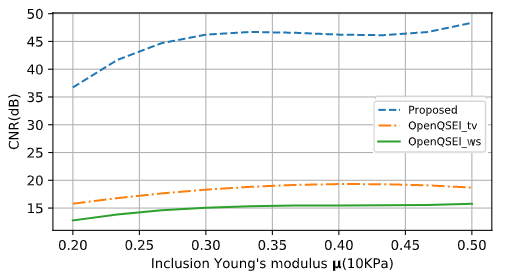}}
\end{minipage}
\vspace{2mm}
\begin{minipage}[b]{0.49\linewidth}
  \centering
  \centerline{\includegraphics[width=4.3cm]{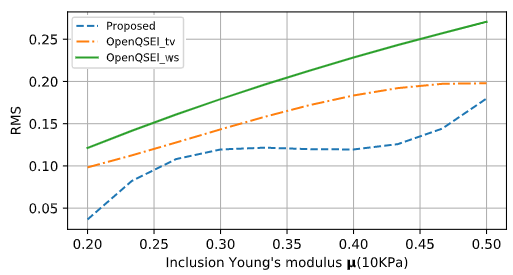}}
\end{minipage}
\caption{
CNR and RMS performance metrics achieved by the proposed approach and OpenQSEI with TV and weighted-smoothness (ws) regularizers for a range of ground-truth inclusion Young's modulus  $\mathbf{E}_{true}$ settings.
   SNR=30dB. }
\label{fig:fig5}
\end{figure}

Next, we compare the performance of our approach with that of OpenQSEI \cite{matlab} with TV and weighted smoothness regularizers \cite{marvin3}.
Reconstruction results for two noise levels are shown in Fig.~\ref{fig:fig2}. OpenQSEI is highly sensitive to noise in estimating the Young's modulus, due to Jacobian matrix computation. Our proposed approach is more robust to noise and estimates the Young's modulus better, as a result of our signal-dependent colored noise modeling perspective and the subsequent optimization process. However noisy displacement measurements cause degradation in the performance of both methods in terms of recovering the geometry of the inclusion.
For quantitative analysis, we compute two performance metrics \cite{marvin2}: contrast-to-noise ratio (CNR) and normalized RMS error. Fig.~\ref{fig:fig3} shows the CNR and RMS error behavior for overall $\Delta=0.1\%-10\%$ for the proposed approach and OpenQSEI with TV and weighted smoothness regularizers. The results in Fig.\ref{fig:fig3} show that our proposed approach outperforms both versions of OpenQSEI. 
To assess the impact of the ratio of the inclusion elasticity modulus to the background modulus (hence the contrast), we show the reconstruction results of both approaches (with TV regularizers) for two settings of the ground truth Young's modulus $\mathbf{E}_{true}$ for the inclusion. The results shown in Fig.\ref{fig:fig4} demonstrate that the proposed approach estimates the Young's modulus better in the low-contrast scenario as well. Although the inclusion shape and geometry are affected, the estimated Young's modulus is similar to the ground truth. This is verified by the computed CNR and RMS errors displayed in Fig.\ref{fig:fig5} for a range of $\mathbf{E}_{true}$. 

\section{Conclusion}
\label{conclude}

In this article, we presented a new statistical framework and an associated computational imaging algorithm for model-based quasi-static ultrasound elastography. The core of our contribution involves the formulation of an integrated optimization problem for estimating the Young's modulus based on noisy displacement and force measurements. This integrated formulation is driven by statistical modeling of the observation process involving signal-dependent colored noise and contains a total variation regularizer for elasticity. We propose a fixed-point iterative algorithm involving the use of proximal splitting methods to solve this optimization problem. Our preliminary results demonstrate the effectiveness and robustness of the proposed statistical modeling and optimization approach. Thanks to the use of proximal splitting methods, more advanced regularizers can easily be incorporated into our proposed framework.

\bibliographystyle{IEEEbib}
\bibliography{Mybib}
\end{document}